\documentclass[%
aip,
rsi,%
amsmath,amssymb,
reprint,%
]{revtex4-1}

\usepackage{graphicx}
\usepackage{dcolumn}
\usepackage{bm}

\begin{document}
	
	\title{(3+1)D-printed adiabatic 1-to-N broadband couplers}
	
	\author{Adrià Grabulosa}
	\affiliation{FEMTO-ST Institute/Optics Department, CNRS \& University Bourgogne Franche-Comt\'e, \\15B avenue des Montboucons,
		Besan\c con Cedex, 25030, France}

	\author{Xavier Porte}%
	\email{javier.portefemto-st.fr}
	\affiliation{FEMTO-ST Institute/Optics Department, CNRS \& University Bourgogne Franche-Comt\'e, 
	\\15B avenue des Montboucons,
	Besan\c con Cedex, 25030, France}%

	\author{Erik Jung}
	\affiliation{FEMTO-ST Institute/Optics Department, CNRS \& University Bourgogne Franche-Comt\'e, \\15B avenue des Montboucons,
	Besan\c con Cedex, 25030, France}

	\author{Johnny Moughames}
	\affiliation{FEMTO-ST Institute/Optics Department, CNRS \& University Bourgogne Franche-Comt\'e, \\15B avenue des Montboucons,
		Besan\c con Cedex, 25030, France}

	\author{Muamer Kadic}
	\affiliation{FEMTO-ST Institute/Optics Department, CNRS \& University Bourgogne Franche-Comt\'e, \\15B avenue des Montboucons,
	Besan\c con Cedex, 25030, France}

	\author{Daniel Brunner}
	\affiliation{FEMTO-ST Institute/Optics Department, CNRS \& University Bourgogne Franche-Comt\'e, \\15B avenue des Montboucons,
		Besan\c con Cedex, 25030, France
	}%

	\date{\today}
	
	\begin{abstract}
		
	We report single-mode 3D optical couplers leveraging adiabatic power transfer towards up to 4 output ports.
	We use the CMOS compatible additive (3+1)D \emph{flash}-TPP printing for fast and scalable fabrication.
	Coupling optical losses of such devices are reduced below $\sim$~0.06~dB by tailoring the coupling and waveguides geometry, and we demonstrate almost octave-spanning broadband functionality from 520~nm to 980~nm.

	\end{abstract}
	
	\maketitle
	
\section*{Introduction}	
	
Low-loss single-mode optical coupling is a fundamental photonic tool, in both, classical and quantum settings.
Adiabatic coupling can achieve highly efficient and broadband single-mode coupling using a tapered/inversely-tapered waveguide sequence~\cite{Son2018}, and it is a widespread technique in current 2D photonic integrated circuits technology~\cite{Marchetti2019,Tsuchizawa2005}.
Optical power transfer between the input and output waveguides is achieved through evanescent coupling, where the optical mode adiabatically leaks from the core of the tapering input waveguides through the cladding into the inversely-tapering output waveguides. Furthermore, this principle has been proposed as an efficient-to-integrate scheme for mode-selective coupling~\cite{Riesen2013a} in multi-spatial modes optical communications.

For advantageous scaling of future photonic networks, unlocking the third dimension for integration is essential~\cite{Moughames2020}.
Here, we experimentally evaluate different tapering strategies in additively (3+1)D-printed~\cite{Porte2021} single-mode couplers with 1 input and up to 4 outputs.
We demonstrate that global losses remain <~2~dB for an exceptionally wide wavelengths range almost spanning an octave from 520~nm to 980~nm, with only $\sim$~0.2~dB at optimal conditions. 
Crucially, our 3D lithography fabrication technology is additive and CMOS compatible.

\begin{figure}[t]
	\centering
	\includegraphics[width=1\linewidth]{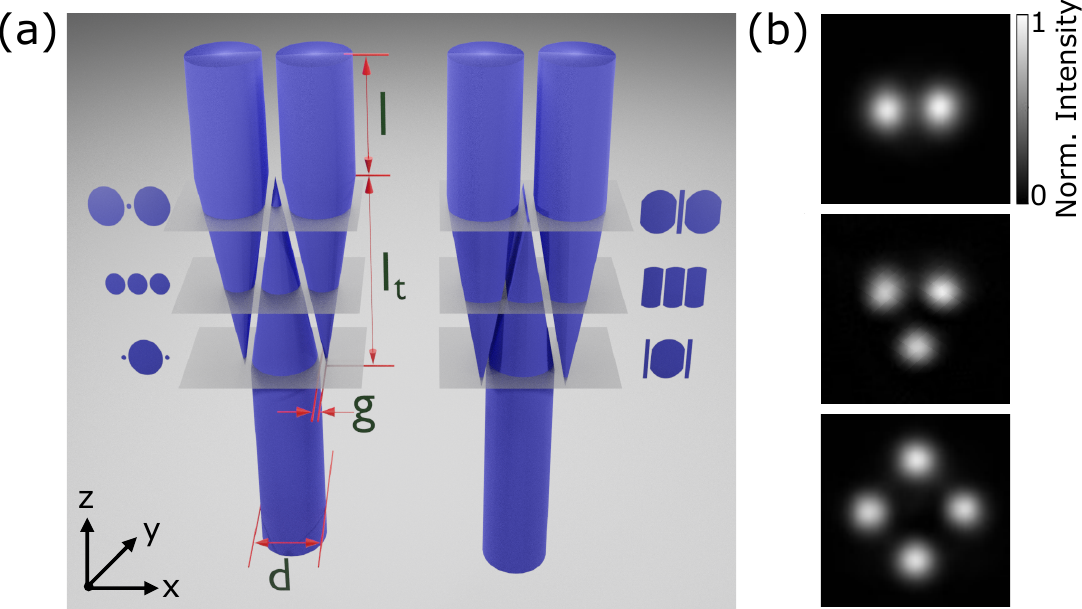}
	\caption{
		(a) Design of the conical (left) and truncated rod (right) geometries of (3+1)D \emph{flash}-TPP printed 1 to 2 couplers leveraging adiabatic power transfer.
		(b) Experimental output intensity profiles of the 1 to 2, 3 and 4 adiabatic couplers with truncated rod geometry. 
	}
	\label{fig:Fig1}
	\vspace{-0.5cm}
\end{figure}

\section*{3D adiabatic couplers}

Using the case of 1 to 2 couplers for illustrating the concept, we tapered waveguide cores according to two different strategies as illustrated in Fig. \ref{fig:Fig1} (a), where the left (right) panel shows conical (truncated rod) geometries, respectively.
In both, the waveguide cross-section continuously changes along the propagation direction $z$ from an input diameter $d$ through a taper-length $l_{\text{t}}$, which is intrinsically linked to the beating length $z_{\text{b}}={\lambda}$/${\Delta{n}}$~\cite{Snyder1983}. 
To all output ports we added a straight section $l=30~\mu$m to minimize cross-talk outside the tapered section.
In the conical geometry, the waveguide cross-section shrinks at an equal rate $d/l_{\text{t}}$ along $(x,y)$.
In the truncated rod geometry, the waveguide only shrinks at that rate along $x$, while along $y$ it retains its original diameter $d$.
The truncated rod design restricts coupling to be parallel to the splitting direction and increases the effective interface-area of the waveguides.
To achieve efficient adiabatic overlapping of optical modes, we inversely tapered input and output waveguides with equal taper-rate and therefore identical symmetry in order to match their effective modal index. 
This approach was used for 1 to 3 and 1 to 4 couplers, too. Figure \ref{fig:Fig1}(b) depicts the single-mode output profile intensities of truncated rod adiabatic couplers with 2 (top), 3 (middle) and 4 (bottom) outputs.    
In the following investigations we consider the truncated rod couplers, which have on average $\sim 3$ times lower overall losses than the conical couplers, presumably due to the extra directionality and increased effective transfer-area. 

\section*{Fabrication}

We leveraged rapid fabrication by combining one- (OPP) and two-photon polymerization (TPP) in the (3+1)D \emph{flash}-TPP lithography concept, saving up to $\approx 90~\%$ of fabrication time~\cite{Grabulosa2022}. 
We use the commercial 3D direct-laser writing Nanoscribe GmbH (Photonics Professional GT) system and the liquid negative-tone IP-S photoresist for the fabrication.
The waveguide cores are printed in a single-step via TPP, with an optimal laser power (LP~=~15~mW) and small hatching distance ($h=0.4~\mu$m), to ensure smooth surfaces.
Mechanical supports, i.e. the cuboid-surface, are printed with larger hatching distance ($h=0.8~\mu$m).
The slicing distance is maximized to $s=1~\mu$m since it does not crucially affect optical performance for purely vertical waveguides. 
After complete-TPP of the IP-S photoresist ($n\approx$ 1.51)~\cite{Schmid2019}, the unexposed photoresist is removed following a standard two-step development process.
Finally, the entire 3D circuit is UV blanket exposed, polymerizing the unexposed volume inside the photonic chip via OPP with a UV exposure dose of 3000~mJ/cm$^2$.
Furthermore, we polymerized with a low TPP laser power (LP~=~1~mW) the volume in-between tapers.
This provides an auxiliary matrix that improves the stability for complex structures during fabrication without significantly modifying the refractive index contrast ($\Delta{n} \approx 5\times10^{-3}$) between core-cladding waveguides~\cite{Grabulosa2022}.

\begin{figure}[t]
	\centering
	\includegraphics[width=1\linewidth]{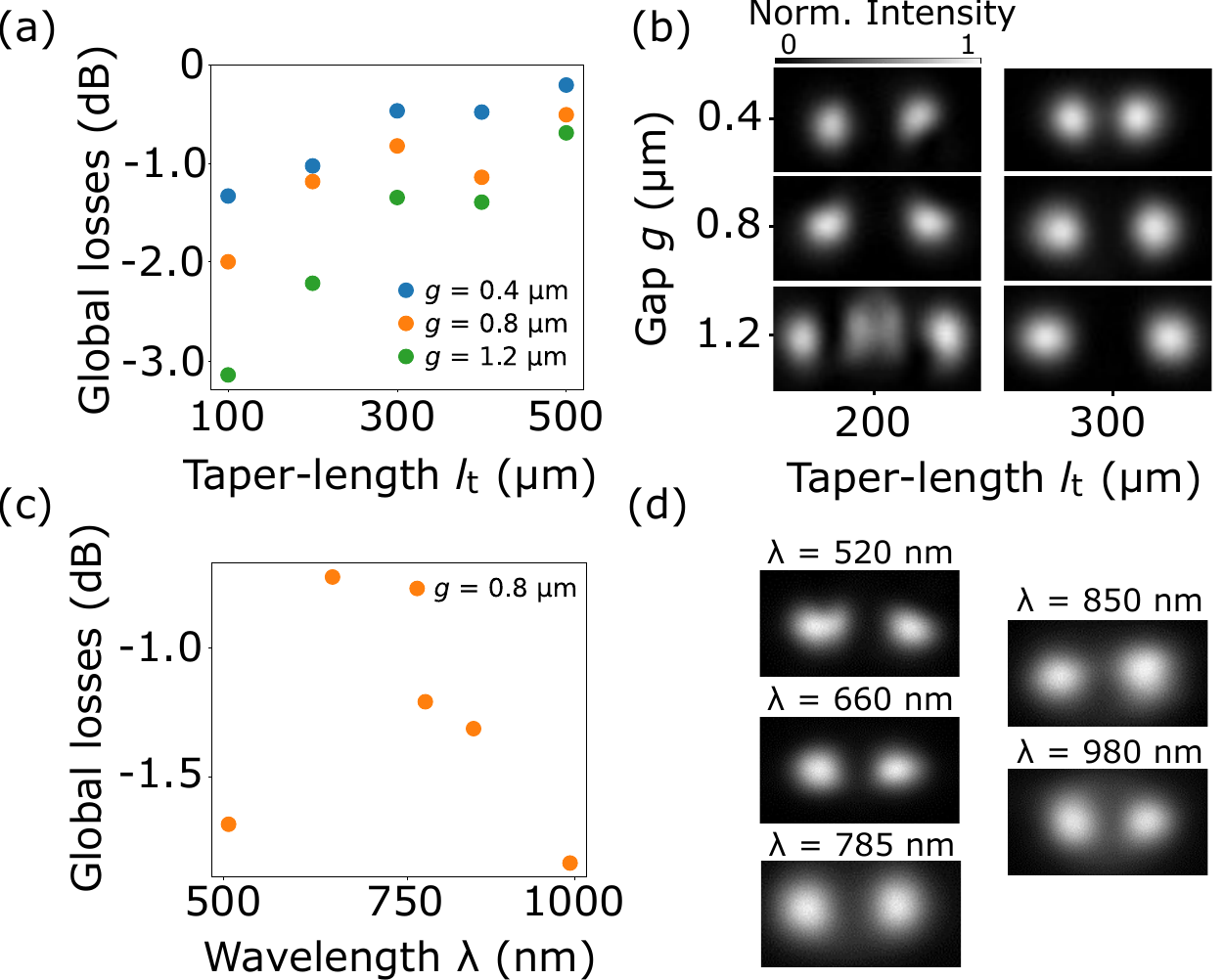}
	\caption{
		(a) Global losses of 1 to 2 couplers with truncated rod geometry for gaps $g\in \{0.4,~0.8,~1.2\}~\mu$m and taper-lengths $l_{\text{t}}\in \{100:100:500\}~\mu$m. 
		(b) Output intensity profiles of the 1 to 2 adiabatic couplers for taper-length $l_{\text{t}} = 200~\mu$m (left) and $l_{\text{t}} = 300~\mu$m (right) and gaps $g\in \{0.4,~0.8,~1.2\}~\mu$m (top to bottom).
		(c) Global losses versus injection light wavelength $\lambda$ of the 1 to 2 adiabatic couplers with $l_{\text{t}} = 500~\mu$m and $g = 0.8~\mu$m. 
		(d) Output intensity profiles from (c) over $\Delta \lambda \sim$~500~nm.
	}
	\label{fig:Fig2}
\end{figure}

\section*{Results}

We evaluate the performance of the couplers via the global losses, which include injection, coupling and propagation losses.
Figure \ref{fig:Fig2}(a) shows the global losses of the 1 to 2 adiabatic couplers with truncated rod geometry for different gaps $g\in~\{0.4,~0.8,~1.2\}~\mu$m, where we scan taper-length $l_{\text{t}}$ from 100 to 500$~\mu$m, and we generally use a diameter of $d=3.3~\mu$m.
We find optimal coupling behavior for $l_{\text{t}} = 500~\mu$m and $g = 0.4~\mu$m, with total losses of $\sim$~0.2~dB, which corresponds to an exceptionally low $\sim$~0.06~dB coupling losses~\cite{Grabulosa2022}, and output intensities at the two output ports differ only by $\sim$ 3.4~$\%$.
From the 1 to 2 output intensity profiles in Fig. \ref{fig:Fig2}(b) it is clear that for $l_{\text{t}} = 200~\mu$m (left) the outputs profiles are not the fundamental LP$_{01}$ mode, and for $g = 1.2~\mu$m (bottom) individual output modes are not sufficiently decoupled.
In contrast, we obtain full splitting of LP$_{01}$ single-modes for $l_{\text{t}} = 300~\mu$m (right) for all $g$.
Consequently, the adiabatic criterion of our 1 to 2 couplers is fulfilled for a taper-length $l_{\text{t}}~$>$~200~\mu$m, which agrees with the theoretical value of $~z_{\text{b}} \approx 132~\mu$m, at injection wavelenght $\lambda = 660~$nm~\cite{Snyder1983}. 
The adiabatic criterion was numerically validated by 2D-simulations via COMSOL Multiphysics.
There, the fundamental eigenmode is launched at the waveguide's input facet via the \textit{Port} boundary conditions, and then propagates throughout a 2D projection of the splitters from Fig. \ref{fig:Fig1}(a) with scattering boundary conditions and $\Delta{n} \approx 5\times10^{-3}$. 
This further confirms the adiabatic signature of our truncated rod 3D optical splitters.
Finally, considering the optimal parameters $l_{\text{t}} = 500~\mu$m and $g = 0.4~\mu$m, we fabricated 1 to 3 and 1 to 4 adiabatic couplers (cf. Fig. \ref{fig:Fig1}(b)) with total losses of only $\sim$~0.7~dB and intensity difference between output ports of $\sim$~4.6~$\%$ ($\sim$~6.1~$\%$) for 1 to 3 (1 to 4) adiabatic couplers.

A major advantage of adiabatic power transfer compared to interference-based directional couplers is a wavelength-independent splitting of optical signals. 
We test the broadband functionality of our 1 to 2 adiabatic couplers with $l_{\text{t}} = 500~\mu$m by injecting different wavelengths ranging from $\lambda$~=~520~nm to $\lambda$~=~980~nm.
Crucially, the bulk absorption of the IP-S photoresist does not play a role on our short relevant length $l_{\text{t}}$ across the entire wavelenght range investigated here~\cite{Schmid2019}.
According to data shown in Fig. \ref{fig:Fig2}(c), global losses remain below $\sim$~2~dB for the 1 to 2 adiabatic couplers over this range almost spanning an octave. 
For $\lambda \geq 660~$nm, global losses start to increase due to lower modal confinement for which larger gap $g$ is required for adiabaticity.
At $\lambda = 520~$nm, we approach the single-mode cut-off wavelength, i.e. $\sim$~480~nm.
Higher-order modes are excited for which the evanescent coupling decreases due to reduced modal overlap. 
Finally, Fig. \ref{fig:Fig2}(d) depicts the output intensity profiles of the couplers across this entire range of wavelengths, which clearly show the discussed effects of higher-order modes as well as non-separated single-modes.

\section*{Conclusions}

In summary, we have shown the (3+1)D \emph{flash}-TPP fabrication of single-mode 3D optical couplers leveraging adiabatic power transfer between one input and up to 4 output ports.
After optimization of the coupling geometry, we obtain $\sim$~0.2~dB ($\sim$~0.06~dB) global (coupling) losses and adiabatic broadband functionality over $\Delta \lambda \sim$~500~nm. The scalability of output ports here addressed~\cite{Moughames2020} can only be achieved by using the three spatial dimensions. 
This further demonstrates this printing strategy as a powerful tool for complex 3D integrated photonic circuits, particularly towards future integration of parallel optical interconnects.

\section*{Funding}
European Union’s Horizon 2020 (713694); Volkswagen Foundation (NeuroQNet II); Agence Nationale de la Recherche (ANR-17-EURE-0002, ANR-15-IDEX-0003); Region Bourgogne Franche-Comté; French Investissements d’Avenir; French RENATECH network and FEMTO-ST technological facility.
	
\section*{References}	
\bibliography{references}
\bibliographystyle{ieeetr}
	
\end{document}